\documentclass{article}
\usepackage[utf8]{inputenc}
\usepackage{graphicx}
\usepackage{epsfig}
\usepackage{amsmath}
\usepackage{amssymb}
\usepackage{braket}
\usepackage{indentfirst}
\usepackage{appendix}

\title{Vacuum radiation from massive scalar field}
\date{}
\author{Yu-Song Cao\footnote{caoyusong15@mails.ucas.ac.cn}$~^{1,2}$}

\begin{document}
\maketitle

\noindent $~^{1}$\small{School of Physical Sciences, University of Chinese Academy of Sciences, Beijing, 100049, China}

\noindent $~^{2}$\small{Beijing National Laboratory for Condensed Matter Physics, Institute of Physics, Chinese Academy of Sciences, Beijing 100190, China}

\begin{abstract}
        The vacuum radiation of a massive scalar field is studied by means of a single moving mirror. The field equation with an arbitrary-shaped mirror moving in $(d+1)$ dimensions is given perturbatively in the non-relativistic limit. Explicit results are obtained for a flat mirror moving in $(1+1)$ dimensions and $(3+1)$ dimensions. The vacuum radiation power and vacuum friction force on the mirror are given in $(1+1)$ dimensions. The intrinsic mass of the field is found to suppress the vacuum radiation. In $(3+1)$ dimensions, the modification of the frequency spectra and angular spectra of emitted particles due to the intrinsic mass are obtained. In the limit of $m\to 0$, we recover the results of the massless field.
\end{abstract}

\section{Introduction}
        As a result of Heisenberg's uncertainty relation, the quantum vacuum field obtains non-vanishing eigen-energy, which is associated to the zero-point fluctuation of infinite harmonic oscillators of different frequencies \cite{Itykson}. There are several observable effects to support the existence of vacuum fluctuation, such as the Casimir effect \cite{Casimir48}. In fact, being a fluctuating medium, the quantum vacuum can exhibit astonishing effects under external perturbations. On the one hand, when placed in non-adiabatic time-varying boundary conditions, the quantum vacuum can emit physical particles and exerts a vacuum friction force on the boundary, known as the dynamical Casimir effect \cite{Moore,Davis,RevofMod}. On the other hand, the absence of the global hyperbolicity of curved spacetime will result in squeezing of the field vacuum and lead to the emission of particles \cite{HR1,HR2}.

        From an experimental point of view, the dynamical Casimir effect was originally proposed in an optical cavity with one moving mirror \cite{Moore,Davis,RevofMod,A57,Dodonov}. Then it was discovered that the vacuum radiation can also occur with one mirror moving in a vacuum \cite{A94,Neto94}. For experimental feasibility, the quantum field is chosen as the electromagnetic field, in which conductors
         can serve as mirrors to impose the required boundary conditions \cite{Moore,Davis,RevofMod,A57,A94,Neto94}. Based on the explicit formulas, it was found that the electromagnetic field is decomposed into components corresponding to the electric field parallel (TM) or perpendicular (TE) to the plane of incidence as a result of gauge symmetry. Moreover, all the dynamical information, i.e., the time-varying parts, are solely contained
          in the magnitude value of the electromagnetic field at each spatial point \cite{A94,Neto94}. The motion equation and boundary condition of the TE mode differs only from those of a massless scalar field with a constant vector at each spatial point, which results in the identical radiation power and vacuum friction force of the TE mode of an electromagnetic field vacuum and a massless scalar field vacuum when the mirrors are travelling through the same trajectory \cite{D82,A94,Neto94,Kardar13}.

        From a theoretical point of view, the dynamical Casimir effect can serve as a counterpart of Hawking radiation and the Unruh effect according to the general equivalence principle \cite{UE,lust,BH1,BH2,BH3,BH4}. Specifically speaking, the Bogoliubov transformation of field modes in the dynamical Casimir effect as a result of the moving mirror is identical to those from the exponential red-shifting near the horizon of a black hole \cite{RevofMod,lust}. Thus, many efforts have been devoted to the dynamical Casimir effect to simulate the physical phenomenon taking place near a black hole \cite{BH1,BH2,BH3,BH4}. In the real world, most particles, fundamental or not, have intrinsic inertial mass. Thus, it is instructive to study the vacuum radiation effect of a massive field. Currently, many efforts are being devoted to studying the dynamical Casimir effect of a massive field. The vacuum friction effect and motion fluctuation of a mirror with an internal degree of freedom on a specified worldline is given in \cite{Unruh92}. The vacuum radiation power, as well as the Euclidean effective action, of massive field coupled to a moving mirror are derived by means of the path integral approach in \cite{D76,D101}. Furthermore, the dynamical Casimir effect of a massive field in the presence of two mirrors is considered in \cite{D91,haro2007}.

        In this paper, we study the emission of particles from a massive scalar field vacuum with a single moving perfect reflective mirror. As an extension of the results obtained for the same physical system \cite{D76,D101}, we obtain the vacuum friction force and the spatial spectra of the emitted particles. Despite the generality and efficiency of the path integral \cite{D76,D91,D101}, we use the canonical quantization method, together with the scattering theory formula, because they can be used to express the vacuum radiation process more clearly, with the pair emission of vacuum radiation demonstrated by means of the negative sea picture \cite{Kardar13}. We find that when the frequency of the driving force is smaller than twice the particle mass, there will be no vacuum radiation at all. As an overall effect, the intrinsic mass is found to suppress the vacuum radiation. The explicit form of vacuum radiation frequency spectra suggests that the vacuum radiation behavior are modified mainly in the low-frequency regime. This is in agreement with intuition that it is harder to excite heavier particles out of vacuum and the dispersion relation of massless particles and massive particles differs more at low frequency regime. In three dimensions, we demonstrate that the the angular spectra of a massive field is narrower than that of a massless field. The corresponding results obtained in the massless case \cite{D82,A94,Neto94,Kardar13} can be regarded as the limit of $m\rightarrow 0$ of our results. This paper is organized as follows. In Section 2, we present the field equation with the arbitrary shape of moving mirror in $(d+1)$ dimensions perturbatively in the non-relativistic limit. The input-output relations of the field in $(1+1)$ dimensions and the quantization of the field are given in Section 4, enabling us to obtain the formula of the vacuum radiation power and the vacuum friction force on the mirror in Section 4. The frequency and angular spectra of the emitted particles in $(3+1)$ dimensions are discussed in Section 5. We present our conclusions and a discussion of our work in Section 6.

        Throughout this paper the units are chosen as $\hbar=c=1$.

\section{General Relations}
        In this section, we mimic the procedure introduced in \cite{D82} to obtain the field equation with one perfect reflective moving mirror perturbatively. The mirror imposes Dirichlet boundary conditions upon the field
        \begin{equation}\label{eq:eq1}
        \phi(x)|_{S}=0,
        \end{equation}
        where $x$ denotes the spacetime coordinate and $S$ denotes the timelike world sheet of the mirror in $(d+1)$-dimensional spacetime.

\begin{figure}
\includegraphics[width=10.5 cm]{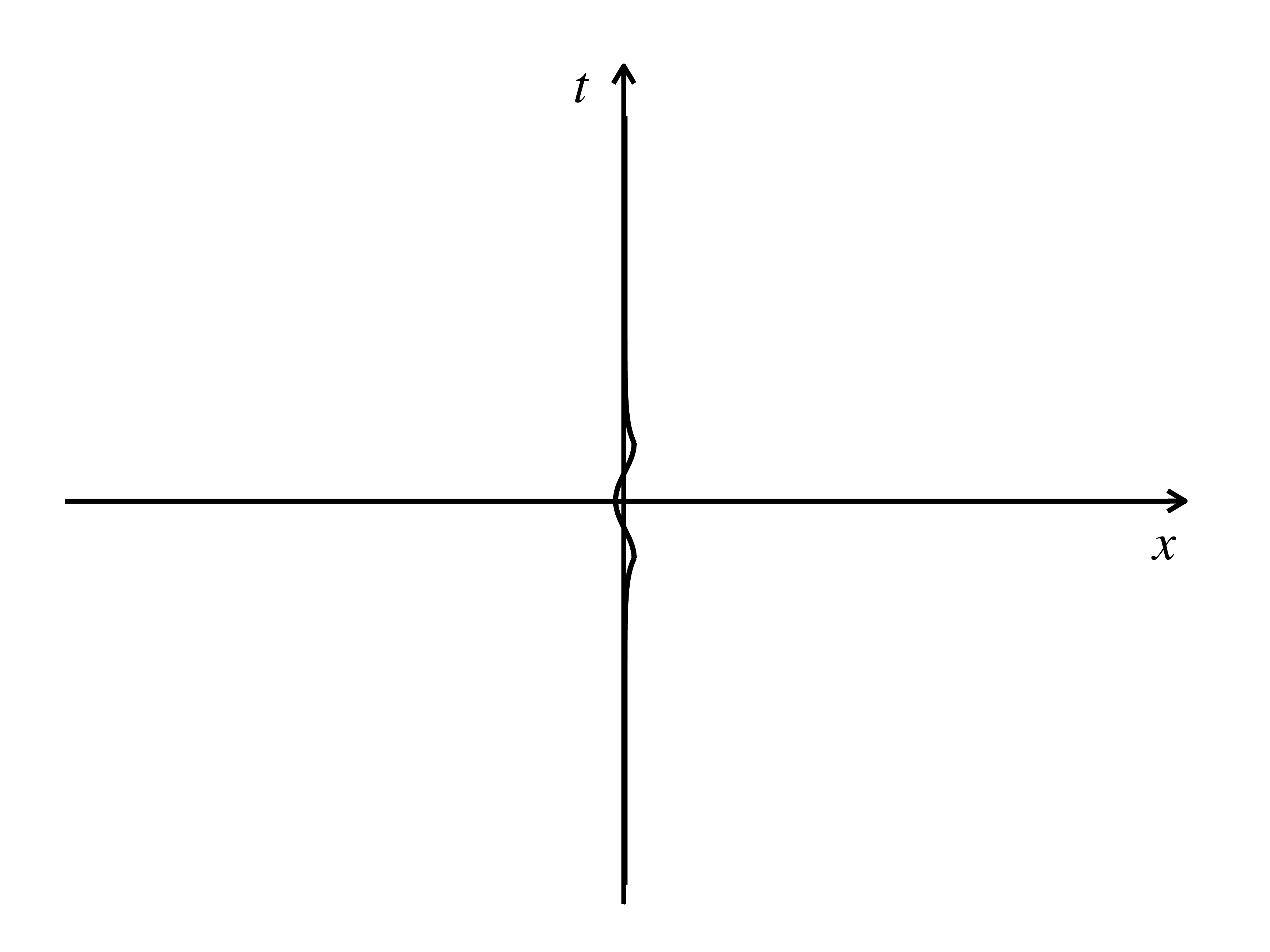}
\caption{A schematic of the mirror's worldline in $(1+1)$-dimensional spacetime, where the mirror approaches a static state at $x=0$ when $t\to\pm\infty$.}
\label{fig:schem}
\end{figure}

        For simplicity, the field equation under consideration is the Klein--Gordon equation
        \begin{equation}
        (\square+m^{2})\phi(x)=0,
        \end{equation}
        where $\square$ is the d'Alembert operator. The asymptotic behavior of the field equation is given as
        \begin{equation}\label{eq:asym}
        \lim_{t\to\infty}\phi(x)=\phi_{0}(x),
        \end{equation}
        where the field $\phi_{0}(x)$ satisfies the following conditions
        \begin{equation}\label{eq:phi0}
        \begin{split}
        &(\square+m^{2})\phi_{0}(x)=0,\\
        &\phi_{0}(x)|_{S_{0}}=0,
        \end{split}
        \end{equation}
        with $S_{0}$ denoting the world sheet of a static mirror. Equation~\eqref{eq:asym} presents an adiabatic approximation of the motion of the mirror, which assumes that the mirror is still when $t\to\pm\infty$. A schematic of the mirror's worldline in $(1+1)$-dimensional spacetime is presented in Figure \ref{fig:schem}.

        The boundary condition in Equation~\eqref{eq:eq1} can be rewritten as
        \begin{equation}\label{eq:eq5}
        \phi[x+\xi(x)]|_{S_{0}}=0,
        \end{equation}
        where $\xi(x)$ is the displacement vector, which translates the points $x$ in $S_{0}$ to $S$. In the following discussion, $\xi(x)$ is considered to be a small quantity.\\
        Expanding Equation~\eqref{eq:eq5} with $\xi(x)$ gives
        \begin{equation}
        [\phi(x)+\xi^{\mu}(x)\partial_{\mu}\phi(x)]|_{S_{0}}=0.
        \end{equation}

        Defining
        \begin{equation}\label{eq:define}
        \phi_{1}=\phi-\phi_{0}
        \end{equation}
        as a scattered field, we then obtain
        \begin{equation}\label{eq:phione}
        \begin{split}
        &(\square+m^{2})\phi_{1}(x)=0,\\
        &\phi_{1}(x)|_{S_{0}}=-\xi^{\mu}(x)\partial_{\mu}\phi_{0}(x)|_{S_{0}},\\
        &\phi_{1}(x)|_{t\to\infty}=0.
        \end{split}
        \end{equation}

        With the help of the generalized Green identity
        \begin{equation}
        \int[\psi(\square+m^{2})\phi-\phi(\square+m^{2})\psi]dV=\int_{\partial V}(\psi\partial^{\mu}\phi-\phi\partial^{\mu}\psi)d\Sigma{\mu}
        \end{equation}
        we derive the solution of $\phi_{1}(x)$ as
        \begin{equation}\label{eq:phi1}
        \begin{split}
        \phi_{1}(x)&=\int_{S_{0}}\phi_{1}(x')\partial_{\mu}'G(x,x')d\Sigma^{\mu'}\\
        &=-\int_{S_{0}}\xi^{\nu}(x')\partial_{\nu}'\phi_{0}(x')\partial_{\mu}'G(x,x')d{\Sigma^{\mu'}},
        \end{split}
        \end{equation}
        where $G(x,x')$ is the retarded Green function under Dirichlet boundary conditions
        \begin{equation}\label{eq:Green}
        \begin{split}
        &(\square+m^{2})G(x,x')=(\square'+m^{2})G(x,x')=\delta(x-x'),\\
        &G(x,x')|_{x\in S_{0}}=G(x,x')|_{x'\in S_{0}}=0,\\
        &G(x,x')=0~~\text{if}~~t<t'.
        \end{split}
        \end{equation}

        Thus, we can express the the scattered field $\phi(x)$ in terms of $\phi_{0}(x)$ and the Green function.

        In the following text, the physical quantity that will be used is the energy-momentum tensor, which is given as
        \begin{equation}
        T_{\mu\nu}=\partial_{\mu}\phi\partial_{\nu}\phi-\frac{1}{2}g_{\mu\nu}[(\partial_{\rho}\phi)^{2}-m^{2}\phi^{2}].
        \end{equation}

        It is notable that the Dirichlet boundary condition forbids the Hamiltonian expression of the field dynamics as for the massless field scenario \cite{Moore,D76}.

\section{Input-Output Relation and Bogoliubov Coefficient}\label{sec3}
        In this section we will obtain the ingredients required to compute the vacuum radiation power and vacuum friction force. To make things simpler, we will work in $(1+1)$ dimensions, in which the mirror becomes a movable perfectly reflective point. As mentioned in \cite{Moore,D76}, there is no Hilbert space for the field dynamics in the presence of a moving mirror. However, the asymptotic condition Equation~\eqref{eq:asym} suggests that it is legitimate to define the Hilbert space of the field at $t\to\pm\infty$  and the field at $t=-\infty$ and $t=\infty$ can be related via the Green functions~\cite{A94,Kardar13,Neto94}. As a consequence of this, the particle creation and annihilation operators at $t=-\infty$ and $t=\infty$ are related by the Bogoliubov transformation~\cite{A94,Kardar13,Neto94}. In this section, we will follow this methodology. Note that from this section onwards, we will treat the spacetime variables in the field equation separately, with $x$ denoting the space coordinate only.

        We set $S_{0}$ to be the worldline $x=0$ around which the mirror undergoes a small displacement. The scattered field Equation~\eqref{eq:phi1} in $(1+1)$ dimensions takes the following form
        \begin{equation}\label{eq:phi11}
        \phi_{1}(t,x)=\int_{x'=0}{dt'}q(t')\partial_{x'}\phi_{0}(t',x')\partial_{x'}G(x,t;x',t'),
        \end{equation}
        where $q(t)$ is the trajectory of the mirror. The non-relativistic limit requires $\dot{q}(t)\ll 1$.

        The partial Fourier transformation in $(1+1)$ dimensions is
        \begin{equation}\label{eq:PF}
        f(t,x)=\int\frac{d\omega}{2\pi}e^{-i\omega{t}}f[\omega,x],
        \end{equation}
        in which parentheses and square brackets are used to distinguish the primitive and Fourier-transformed functions.

        Partially Fourier-transforming Equation~\eqref{eq:phi11} gives
        \begin{equation}
        \phi_{1}[\omega,x]=-\phi_{1}[\omega,0]\partial_{x'}G[\omega;x,x'=0],
        \end{equation}
        where we used the fact the Green function is invariant under time translation
        \begin{equation}
        G(t,x;t',x')=G(t-t';x,,x').
        \end{equation}

        After some algebra, the field function can be expressed in terms of the Green function as
        \begin{equation}
        \phi[\omega,x]=\phi_{in}[\omega,x]-\partial_{x'}G_{R}[\omega;x,x'=0]\phi_{1}[\omega,x'=0],
        \end{equation}
        where $\phi_{in}$ denotes the incoming field at $t=-\infty$, satisfying Equation~\eqref{eq:phi0}, whereas the similar relation between $\phi$ and $\phi_{out}$ (field equation at $t=\infty$) can be obtained with the advanced Green function. Thus, we can relate the incoming and outgoing field with the retarded and advanced Green functions:
        \begin{equation}\label{eq:inout}
        \phi_{out}[\omega,x]=\phi_{in}[\omega,x]-(\partial_{x'}G_{R}[\omega;x,x'=0]-\partial_{x'}G_{A}[\omega;x,x'=0])\phi_{1}[\omega,x].
        \end{equation}

        Now our task will be computing the Green functions $G_{R}[\omega;x,x']$ and $G_{A}[\omega;x,x']$. The boundary condition of Equation~\eqref{eq:Green} hints that the retarded Green function $G(t,x;t',x')$ can be derived from the Green function $G^{(0)}(t,x;t',x')$ in free space, together with mirror symmetry. The partially Fourier-transformed retarded Green function in free space can be obtained from
        \begin{equation}
        \begin{split}
        G_{R}^{(0)}(t,x;t',x')&=-\frac{1}{(2\pi)^{2}}\int{d\omega}{dk}\frac{1}{(\omega+i\epsilon)^{2}-k^{2}-m^{2}}e^{-i{\omega}(t-t')+ik(x-x')}\\
        &=-i\int\frac{d\omega}{2\sqrt{\omega^{2}-m^{2}}}e^{-i{\omega}(t-t')+i\sqrt{\omega^{2}-m^{2}}(x-x')},
        \end{split}
        \end{equation}
        whereas the formula of the advanced Green function can be obtained similarly
        \begin{equation}
        \begin{split}
        G_{A}^{(0)}(t,x;t',x')&=-\frac{1}{(2\pi)^{2}}\int{d\omega}{dk}\frac{1}{(\omega-i\epsilon)^{2}-k^{2}-m^{2}}e^{-i{\omega}(t-t')t+ik(x-x')}\\
        &=i\int\frac{d\omega}{2\sqrt{\omega^{2}-m^{2}}}e^{-i{\omega}(t-t')+i\sqrt{\omega^{2}-m^{2}}(x-x')}.
        \end{split}
        \end{equation}

        With the help of mirror symmetry, the Green function under Dirichlet boundary conditions and the Green function in free space are related by means of the following relation
        \begin{equation}
        G(t,x;t',x')=G^{(0)}(t,x;t',x')-G^{(0)}(t,x;t',\bar{x'})
        \end{equation}
        with $\bar{x'}$ denoting the image point of $x'$. In this case $\bar{x'}=-x'$. Note that this formula is valid for both retarded and advanced Green functions. Then, the partially Fourier-transformed Green functions are
        \begin{equation}
        \begin{split}
        &G_{R}[\omega;x,x']=-\frac{i}{2\sqrt{\omega^{2}-m^{2}}}(e^{i\sqrt{\omega^{2}-m^{2}}(x-x')}-e^{-i\sqrt{\omega^{2}-m^{2}}(x+x')}),\\
        &G_{A}[\omega;x,x']=\frac{i}{2\sqrt{\omega^{2}-m^{2}}}(e^{-i\sqrt{\omega^{2}-m^{2}}(x-x')}-e^{i\sqrt{\omega^{2}-m^{2}}(x+x')}),
        \end{split}
        \end{equation}
        from which we get
        \begin{equation}\label{eq:RA}
        \begin{split}
        &\partial_{x'}G_{R}(\omega;x,x'=0)=-e^{i\sqrt{\omega^{2}-m^{2}}x},\\
        &\partial_{x'}G_{A}(\omega;x,x'=0)=-e^{-i\sqrt{\omega^{2}-m^{2}}x}.
        \end{split}
        \end{equation}

        The boundary condition of $\phi_{1}$ given in Equation~\eqref{eq:phione} under the partial Fourier transformation becomes
        \begin{equation}\label{eq:boundary1}
        \phi_{1}[\omega,0]=-\int\frac{d\omega'}{2\pi}q[\omega-\omega']\partial_{x}\phi_{0}[\omega',0].
        \end{equation}

        Substituting Equations~\eqref{eq:RA} and \eqref{eq:boundary1} into Equation~\eqref{eq:inout}, we obtain the input-output relation
        \begin{equation}\label{eq:inputoutput}
        \phi_{out}[\omega,x]=\phi_{in}[\omega,x]-2i\sin(\sqrt{\omega^{2}-m^{2}}x)\int\frac{d\omega'}{2\pi}q[\omega-\omega']\partial_{x}\phi_{in}[\omega',x=0].
        \end{equation}

        At $t\to\pm\infty$, the field can be quantized in a straightforward manner via mode expansion. By the way, the electromagnetic field, as mentioned in Section 1, can be decomposed to TE and TM components as a result of gauge symmetry. For each component, they can be quantized directly via mode expansion \cite{Neto94,A94}. Thus, the difference in quantization procedures doe gauge fields and scalar fields observed in the textbooks of quantum field theory does not take place here \cite{Itykson}.

        In Equation~\eqref{eq:phi0}, the field confined in the left (right) half space is the superposition of the right (left)-moving wave and the left (right)-moving reflected wave, with equal amplitudes for each mode. In the right half space, the field can be mode-expanded as
        \begin{equation}\label{eq:onemode}
        \phi_{0}(t,x)=i\int_{0}^{\infty}\frac{dk}{2\pi}\sqrt{\frac{2}{\omega_{k}}}\sin(kx)[e^{-i{\omega_{k}}t}a_{k}-e^{i{\omega_{k}}t}a^{\dagger}_{k}]
        \end{equation}
        with the dispersion relation $\omega_{k}=\sqrt{k^{2}+m^{2}}$. The mode number $k$ takes non-negative values due to Equation~\eqref{eq:onemode}, being the standing wave. Note that both the positive frequency part and the negative frequency part are present in Equation~\eqref{eq:onemode}. The quantization procedure takes the amplitudes of the positive frequency modes and negative frequency modes into operators $a_{k}$ and $a^{\dagger}_{k}$, satisfying
        \begin{equation}
        [a_{k},a^{\dagger}_{k'}]=2\pi\delta(k-k').
        \end{equation}

        To verify this, we compute the Hamiltonian in the right half space
        \begin{equation}
        \begin{split}
        H&=\int_{x>0}dxT^{00}(t,x)\\
        &=\int_{k>0}\omega_{k}a_{k}^{\dagger}a_{k},
        \end{split}
        \end{equation}
        which is intuitively consistent. Furthermore, the Hamiltonian in the left half space can be derived similarly. The partially Fourier-transformed Equation~\eqref{eq:onemode} gives
        \begin{equation}
        \phi_{in}[\omega,x]=i\sqrt{\frac{2|\omega|}{\omega^{2}-m^{2}}}\sin(\sqrt{\omega^{2}-m^{2}}x)[\theta(\omega)a_{k}^{(in)}-\theta(-\omega)a_{-k}^{\dagger(in)}].
        \end{equation}

        A similar formula can be obtained for $\phi_{out}$. Together with the input-output relation presented in Equation~\eqref{eq:inputoutput}, we can obtain the Bogoliubov coefficient between the annihilation operators of incoming and outgoing field:
        \begin{equation}\label{eq:Bogoliubov}
        a^{(out)}_{k}=a^{(in)}_{k}+2i\frac{k}{\sqrt{\omega}}\int\frac{d\omega'}{2\pi}q[\omega-\omega']\sqrt{|\omega'|}[\theta(\omega')a^{(in)}_{k'}-\theta(-\omega')a^{\dagger(in)}_{-k'}].
        \end{equation}

        Equation~\eqref{eq:Bogoliubov} presents a clear demonstration of how particles are generated from the vacuum with the negative sea picture, in which the vacuum is regarded as a state where all the negative energy bands are occupied by unobservable particles. Note that this picture is artificial and we need not be bothered with the boson statistics. The motion of the mirror excites one negative energy particle to a positive band, thus making it observable and leaving a hole in the negative sea. In the eyes of an observer, the hole in the negative sea behaves like a particle with the opposite energy and momentum to the missing particle, which can be inferred from Equation~\eqref{eq:Bogoliubov}. This means the particles are always created in pairs. The dispersion relation of the massive scalar field $\omega_{k}^{2}=k^{2}+m^{2}$ shows that there exists a $2m$ gap between the positive and negative energy bands.  Equation~\eqref{eq:Bogoliubov} indicates that only the mirror frequency $\Omega$, satisfying $\Omega>2m$, can generate particles out of a vacuum, which is also verified via the path integral approach in \cite{D76}.

\section{Vacuum Radiation and Vacuum Friction}\label{sec4}
        In the earlier works \cite{A94,Neto94}, with Bogoliubov coefficients, the vacuum radiation power and the corresponding vacuum friction force were computed for the specified motion of the mirror. However, there is actually more that can be accomplished in this area. The Bogoliubov coefficients in \cite{Kardar13}, interpreted as $S$-matrix elements, can give rise to the vacuum radiation power and the vacuum friction force for the arbitrary trajectory of the mirror. In this section, we will adopt this technique for the massive field.

        Before diving into the details, we first examine the scattered field. Equations~\eqref{eq:inputoutput} and \eqref{eq:onemode} demonstrate that the incoming field is scattered into normal reflective waves as well as sidebands. Based on Equation~\eqref{eq:Bogoliubov}, the sideband of an incoming wave with frequency $\omega$ is given by $\pm(\omega-\Omega)$, where $\Omega$ is the frequency of the mirror. Equation~\eqref{eq:onemode} indicates that the reflective wave is absorbed in $\phi_{0}$; thus, only the sideband is responsible for the vacuum radiation. The $S$-matrix is defined as \cite{Kardar13}
        $$a_{k}^{(out)}=\int dk'S(k,k')a_{k'}^{(in)}$$
        of which the non-vanishing off-diagonal element, in the frequency representation
        \begin{equation}
        S_{\omega+\Omega,\omega}=2i\sqrt{\frac{(\omega+\Omega)^{2}-m^{2}}{\omega+\Omega}}\sqrt{|\omega|}q[\Omega]
        \end{equation}
        is responsible for vacuum radiation. For a massive field, there exists a $2m$ gap in the mirror's frequency spectra, as suggested by Equation~\eqref{eq:Bogoliubov}. So the radiated power
        \begin{equation}\label{eq:energy}
        \begin{split}
        P&=2\int_{2m}^{+\infty}\frac{d\Omega}{2\pi}\int_{-\Omega}^{-m}\frac{d\omega}{2\pi}(\omega+\Omega)|S_{\omega+\Omega,\omega}|^{2}\\
        &=8\int_{2m}^{+\infty}\frac{d\Omega}{2\pi}|q[\Omega]|^{2}\int_{m-\Omega}^{-m}\frac{d\omega}{2\pi}[(\omega+\Omega)^{2}-m^{2}]|\omega|\\
        &=\frac{4}{\pi}\int_{2m}^{+\infty}\frac{d\Omega}{2\pi}|q[\Omega]|^{2}(\frac{1}{12}\Omega^{4}-m^{2}\Omega^{2}+\frac{4}{3}m^{3}\Omega),
        \end{split}
        \end{equation}
        which goes back to the result of the massless field in the limit of $m{\rightarrow}0$ and a factor $2$ is adapted to account for scattering from both sides. Adopting the fluctuation-dispassion theorem, the radiated power is related to the vacuum friction force by
        \begin{equation}
        P=\int dtF(t)\dot{q}(t),
        \end{equation}
        where $F$ is the vacuum friction force. The fluctuation-dispassion theorem under partial Fourier transformation becomes
        $${\int}dtF(t)\dot{q}(t)=\frac{i}{2\pi}{\int}d\omega{\omega}F[\omega]q^{*}[\omega].$$

        Together with Equation~\eqref{eq:energy}, the Fourier component of the vacuum friction force is
        \begin{equation}\label{eq:fomega}
        F[\Omega]=-\frac{4i}{\pi}q[\Omega]({\frac{1}{12}\Omega^{3}-m^{2}\Omega+\frac{4}{3}m^{3}}).
        \end{equation}

\begin{figure}
\includegraphics[width=10.5 cm]{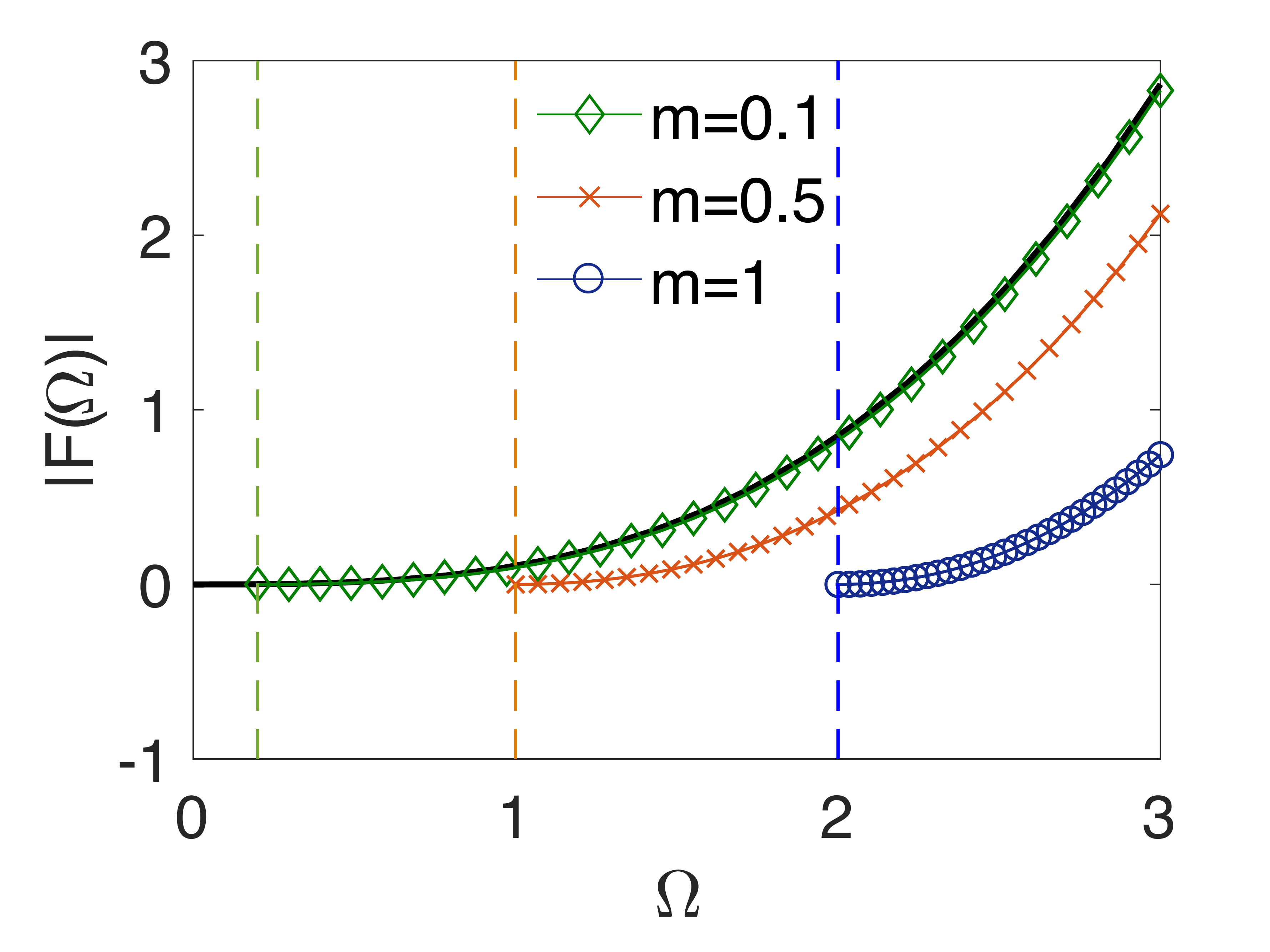}
\caption{Frequency spectra of $|F[\Omega]|$, in which we set the Fourier components of motion $q[\Omega]=1$. The black solid line denotes the results for the massless field. Note the units introduced in Section 1; the dimension of all the quantities can be expressed in terms of mass throughout this paper.}
\label{fig:f}
\end{figure}

        Fourier-transforming Equation~\eqref{eq:fomega} gives the vacuum friction force in the time domain as
        \begin{equation}\label{eq:force}
        F(t)=\int_{-\infty}^{+\infty}\frac{d\omega}{2\pi}\theta(\omega^{2}-4m^{2})F[\omega]e^{-i{\omega}t},
        \end{equation}
        Note the theta function is present due to the frequency gap given in \mbox{Equation~\eqref{eq:energy}.}  This is intuitively consistent as no vacuum friction force would take place if no particles were emitted. In Figure~\ref{fig:f} we can see that the mass term suppresses the vacuum radiation by expanding the frequency gap $2m$ and suppress the vacuum radiation power for given driven frequency $\Omega$. The upper bound and lower bound of the second row in Equations~\eqref{eq:energy} shows the window $-m<-\omega<m-\Omega$ of the negative energy band will contribute, which can be inferred from the negative sea picture. Figure~\ref{fig:f} also demonstrates the mass term mainly influences the low-frequency regime of the vacuum radiation. In the high-frequency regime, Equations~\eqref{eq:energy} and \eqref{eq:fomega} indicate that the behavior of vacuum radiation is influenced little by the particle mass. In contrast with the vacuum radiation of a massless field, Equation~\eqref{eq:force} tells us that the backaction force is no longer determined by $q^{(3)}(t)$ (the third derivative of time) but rather in a complicated form for the massive field. In contrast with the massless cases which can be computed with conformal symmetry, the results obtained in this paper are only valid for small $q(t)$ and $\dot{q}(t)\ll 1$.

\section{Spatial Spectra of Emitted Particles}\label{sec5}
        In this section, we consider the spatial distribution of the radiated particles. This means we are going to work in $(3+1)$ dimensions. In \cite{A94}, the spatial spectra of vacuum radiation of the electromagnetic field are derived using the Bogoliubov coefficient. In the following section, we will adapt this methodologyIn $(3+1)$ dimensions, in which the mirror is an infinitely large plane oscillating around $x=0$, parallel to the $y-z$ plane.

        Firstly, we give the partially Fourier-transformed technique in three dimensions~\cite{A94,Neto94}
        \begin{equation}
        F[x,\boldsymbol{k}_{\parallel},\omega]=\frac{1}{S}\int{dt}\int{d^{2}r_{\parallel}}e^{i{\omega}t-i\boldsymbol{k}_{\parallel}\cdot\boldsymbol{r}_{\parallel}}F(x,\boldsymbol{r}_{\parallel},t),
        \end{equation}
        where we take the periodic boundary condition on the $y-z$ plane, which is cut into cells with area $S$, and $\boldsymbol{r}_{\parallel}$ and $\boldsymbol{k}_{\parallel}$ are the position vectors parallel to the $y-z$ plane, respectively. Here we also use parentheses and square brackets to distinguish between the function in the frequency domain and in the time domain. The boundary condition Equation~\eqref{eq:boundary1} in this case under the partial Fourier transformation in the non-relativistic limit gives
        \begin{equation}
        \phi_{1}[0,\boldsymbol{k}_{\parallel},\omega]=-\int\frac{d\omega'}{2\pi}\theta({k'_{x}}^{2})q[\omega-\omega']\partial_{x}\phi_{0}[0,\boldsymbol{k}_{\parallel},\omega],
        \end{equation}
        where the mode number along the $x$-direction is denoted as $k_{x}$. The input-output relation is then obtained using the Green function as
        \begin{equation}\label{eq:3inout}
        \begin{split}
        \phi_{out}[x,\boldsymbol{k}_{\parallel},\omega]&=\phi_{in}[x,\boldsymbol{k}_{\parallel},\omega][\partial_{x'}G_{R}(x,x'=0)-\partial_{x'}G_{A}(x,x'=0)]\phi_{1}[x'=0,\boldsymbol{k}_{\parallel},\omega]\\
        &=\phi_{in}[x,\boldsymbol{k}_{\parallel},\omega]-2i\sin(k_{x}x)\int\frac{d\omega'}{2\pi}\theta({k'_{x}}^{2})q[\omega-\omega']\partial_{x}\phi_{0}[0,\boldsymbol{k}_{\parallel},\omega'].
        \end{split}
        \end{equation}

        Then we quantize the $\phi_{0}$, mode-expanding the field as
        \begin{equation}\label{eq:threemode}
        \phi_{0}(x,\boldsymbol{r}_{\parallel},t)=i\sum_{n}\int_{0}^{+\infty}\frac{dk_{x}}{2\pi}\sqrt{\frac{2}{\omega_{n}S}}\sin(k_{x}x)[e^{-i\omega_{n}t+i\boldsymbol{k}_{\parallel,n}\cdot\boldsymbol{r}_{\parallel}}a_{n}(k_{x})-e^{i\omega_{n}t-i\boldsymbol{k}_{\parallel,n}\cdot\boldsymbol{r}_{\parallel}}a_{n}^{\dagger}(k_{x})],
        \end{equation}
        where $\boldsymbol{k}_{\parallel,n}$ denotes a momentum vector
         parallel to the $y-z$ plane, with sub-indices $n=(n_{y},n_{z})$ marking the quantum number of momentum along the $y$- and $z$-axes. The periodic boundary condition restricts the momentum quantum number $n_{y},n_{z}$ into discrete non-negative integers as follows:
        \begin{equation}
        \boldsymbol{k}_{\parallel,n}=\frac{2\pi}{\sqrt{S}}(n_{y}\boldsymbol{\hat{y}}+n_{z}\boldsymbol{\hat{z}}),
        \end{equation}
        where $\hat{y}$ and $\hat{z}$ are the unit vectors along the $y$- and $z$-axes, respectively. $\boldsymbol{k}_{\parallel,n}$ becomes continuous when taking $S\rightarrow\infty$. Partially Fourier-transforming Equation~\eqref{eq:threemode} gives
        \begin{equation}
        \phi_{0,n}[x,\omega]=i\theta(k_{x}^{2})\sqrt{\frac{2|\omega|}{k_{x}^{2}S}}\sin(k_{x}x)[\theta(\omega)a_{n}(k_{x})-\theta(-\omega)a_{-n}^{\dagger}(-k_{x})].
        \end{equation}

        Together with Equation~\eqref{eq:3inout}, we obtain the Bogoliubov coefficients
        \begin{equation}\label{eq:3Bogoliubov}
        \begin{split}
            a_{out,n}(k_{x})=&a_{in,n}(k_{x})+2i\frac{k_{x}}{\sqrt{\omega_{n}}}\int\frac{d\omega'}{2\pi}\theta({k'_{x}}^{2})\sqrt{|\omega'|}q(\omega_{n}-\omega')[\theta(\omega')a_{in,n}(k'_{x})\\
            &-\theta(-\omega')a_{in,-n}^{\dagger}(-k'_{x})].
        \end{split}
        \end{equation}

        From Equation~\eqref{eq:3Bogoliubov}, we can see that the quantum number $n$ is invariant under the scattering process, whereas the quantum number $k_{x}$ exhibits similar behavior to that observed in Equation~\eqref{eq:Bogoliubov} with a normal reflective wave and sidebands when scattered by the mirror.

        Since the field equation in the Heisenberg picture is identical to the one obtained by means of the variation method with the Lagrangian, the spectra emitted from the vacuum can be derived from the expectation value of the "out" particle number operator over the vacuum's "in" state
        \begin{equation}\label{eq:emitspec}
         \langle 0,in|a_{out,n}^{\dagger}(k_{x})a_{out,n}(k_{x})|0,in\rangle\frac{dk_{x}}{2\pi}.
        \end{equation}

        To grasp the key factor and avoid complexity, the motion of the mirror is considered to be a damped sinusoidal motion
        \begin{equation}
        q(t)=q_{0}e^{-\frac{|t|}{T}}\cos(\omega_{0}t),
        \end{equation}
        where $T$ is a large time factor and $\omega_{0}q_{0}\ll 1$ is required by the non-relativistic limit. In the negative energy sea picture, the mirror's motion generates a positive frequency sideband $\omega_{n}$ from the negative input frequency $\omega'$ with
        \begin{equation}\label{eq:sideband}
        \omega_{n}=\omega_{0}+\omega'.
        \end{equation}

        This also indicates that for the frequencies $\omega_{0}<2m$, no particles can be generated out of the vacuum and that the positive energy sideband frequency $\omega_{n}$ is always smaller than the mirror's frequency $\omega_{0}$. Together with the dispersion relation
        \begin{equation}\label{eq:dispersion}
        \begin{split}
        &\omega'^{2}={k'_{x}}^{2}+\boldsymbol{k}_{\parallel,n}^{2}+m^{2},\\
        &k'_{x}=\sqrt{(\omega_{n}-\omega_{0})^{2}-m^{2}-\boldsymbol{k}_{\parallel,n}^{2}}.
        \end{split}
        \end{equation}
        the number of the particles emitted in the direction $\boldsymbol{k}=(k_{x},\boldsymbol{k}_{\parallel})$ is obtained as

        \begin{equation}\label{eq:emitnum}
         \langle 0,in|a_{out,n}^{\dagger}(k_{x})a_{out,n}(k_{x})|0,in\rangle =q_{0}^{2}\frac{T}{\omega_{n}}(\omega_{n}^{2}-m^{2})(1-\frac{\boldsymbol{k}_{\parallel,n}^{2}}{\omega_{n}^{2}-m^{2}})\sqrt{(\omega_{n}-\omega_{0})^{2}-\boldsymbol{k}_{\parallel,n}^{2}-m^{2}},
        \end{equation}
        which diverges when taking $T\rightarrow \infty$. However, this actually introduces no inconsistency because the subject of interest is the emitting rate, which is obtained via Equation~\eqref{eq:emitnum} divided by $T$, thus cancelling out the infinity \cite{Neto94}.

        The emitted particles are also restrained by the translational invariance parallel to the mirror. This serves as another reason for the particles to have to appear in pairs if they exhibit the non-vanishing total momentum $\boldsymbol{k}_{\parallel}$. Based on Equation~\eqref{eq:sideband}, the positive energy sideband and hole give the energy of the emitted particles:
        \begin{equation}\label{eq:energyconserv}
        \omega_{0}=\omega_{1}+\omega_{2},
        \end{equation}
        where $\omega_{1}$ and $\omega_{2}$ are the energies of the radiated particles, respectively. The plane symmetry parallel to the mirror also requires the total momentum parallel to the $y-z$ plane to be zero; thus,
        \begin{equation}\label{eq:planesym}
        (\omega_{1}^{2}-m^{2})\sin^{2}\theta_{1}=(\omega_{2}^{2}-m^{2})\sin^{2}\theta_{2},
        \end{equation}
        where $\theta_{1}$ and $\theta_{2}$ are the angles between the $x$-axis and the emitted particles\, with the artificially required condition $\theta_{2}>\theta_{1}$. In this sense, $\theta_{2}$ takes a value from $0$ to $\frac{\pi}{2}$.
        Defining the angle between $x$ direction and the emitted particle as $\theta$, then
        \begin{equation}
        \sin\theta=\frac{k_{\parallel,n}}{k}= \frac{k_{\parallel,n}}{\sqrt{\omega_{n}^{2}-m^{2}}}.
        \end{equation}

        The dispersion relation puts a restraint on the input frequency
        \begin{equation}
        -\omega'\geq\sqrt{\boldsymbol{k}_{\parallel,n}^{2}+m^{2}},
        \end{equation}
        with which we can obtain the upper bound for $\theta_{1}$
        \begin{equation}\label{eq:theta0}
        \theta_{1}<\theta_{0}=\arcsin{\sqrt{\frac{(\omega_{0}-\omega_{n})^{2}-m^{2}}{\omega_{n}^{2}-m^{2}}}}.
        \end{equation}

\begin{figure}
\includegraphics[width=10.5 cm]{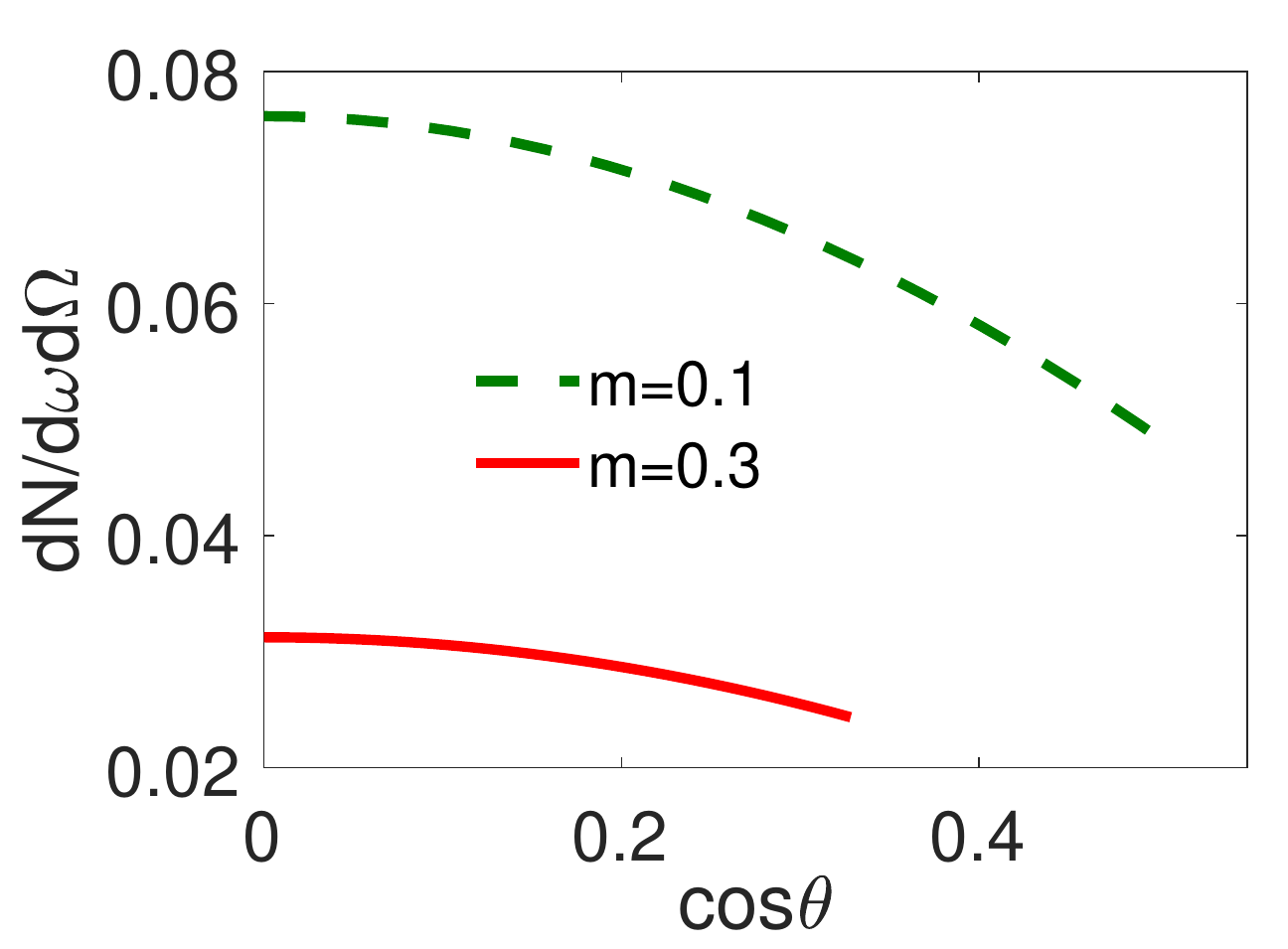}
\caption{The angular spectra of the vacuum radiation, with the prefactor $\frac{TS}{(2\pi)^{3}}$ and driving amplitude $q_{0}$ set to be $1$.}
\label{fig:angular}
\end{figure}

        Equation~\eqref{eq:theta0} restricts the positive frequency sideband $\omega_{n}$ within
        \begin{equation}
        m<\omega_{n}<\omega_{0}-m,
        \end{equation}
        where the first inequality comes from the dispersion relation, in Equation~\eqref{eq:dispersion}, whereas the second comes from the pair emission. The same inequalities can be derived for $-\omega'$. Furthermore, it is easy to see from Figure~\ref{fig:angular} that, as $m$ increases, the particles with frequency $\omega_{1}$ scatter into a narrower solid angle. Figure~\ref{fig:angular} also shows that the mass of the particles will suppress the vacuum radiation as an overall effect.

        Based on Equation~\eqref{eq:emitspec}, the rate of particle emission per frequency and per solid angle is given by
        \begin{equation}
        \frac{dN}{d{\omega}d{\Omega}}(\omega,\theta)=S\frac{\omega^{2}}{(2\pi)^{3}}\langle 0,in|a_{out,n}^{\dagger}(k_{x})a_{out,n}(k_{x})|0,in \rangle.
        \end{equation}

        Since the particles are created in pairs, Equations~\eqref{eq:dispersion} \eqref{eq:energyconserv} and \eqref{eq:planesym} establish a one-to-one correspondence between the motion parameters $\omega_{1},\theta_{1}$, and $\omega_{2},\theta_{2}$. Thus, we can express the angular distribution of the emitted particle in terms of $\omega_{1},\theta_{1}$:
        \begin{equation}\label{eq:angulardistri}
        \frac{dN}{d{\omega}d{\Omega}}(\omega_{1},\theta_{1})=\frac{TS}{(2\pi)^{3}}\frac{q_{0}^{2}}{\omega_{1}}(\omega_{1}^{2}-m^{2})^{2}\sqrt{\omega_{2}^{2}-m^{2}}\cos^{2}\theta_{1}\cos\theta_{2}.
        \end{equation}

        This goes back to the formula of the massless field with $m\to 0$. One must be careful with the variables in Equation~\eqref{eq:angulardistri}, where $\omega_{2},\theta_{2}$ should be taken as functions of $\omega_{1},\theta_{1}$.

        Integrating over the solid angle in Equation~\eqref{eq:angulardistri}, we can obtain the frequency spectra
        \begin{equation}
        \begin{split}
        \frac{dN}{d\omega}(\omega)&=-\frac{TSq_{0}^{2}}{4\pi^{2}}(\omega^{2}-m^{2})[\frac{[(\omega_{0}-\omega)^{2}-m^{2}]^{\frac{3}{2}}}{4\omega}-\frac{\omega_{0}(\omega_{0}-2\omega)\sqrt{(\omega_{0}-\omega)^{2}-m^{2}}}{8\omega}\\
        &-\frac{\omega_{0}^{2}(\omega_{0}-2\omega)^{2}}{8\omega\sqrt{\omega^{2}-m^{2}}}\ln\frac{\sqrt{\omega^{2}-m^{2}}+\sqrt{(\omega_{0}-\omega)^{2}-m^{2}}}{\sqrt{\omega_{0}(\omega_{0}-\omega)}}].
        \end{split}
        \end{equation}

\begin{figure}
\includegraphics[width=10.5 cm]{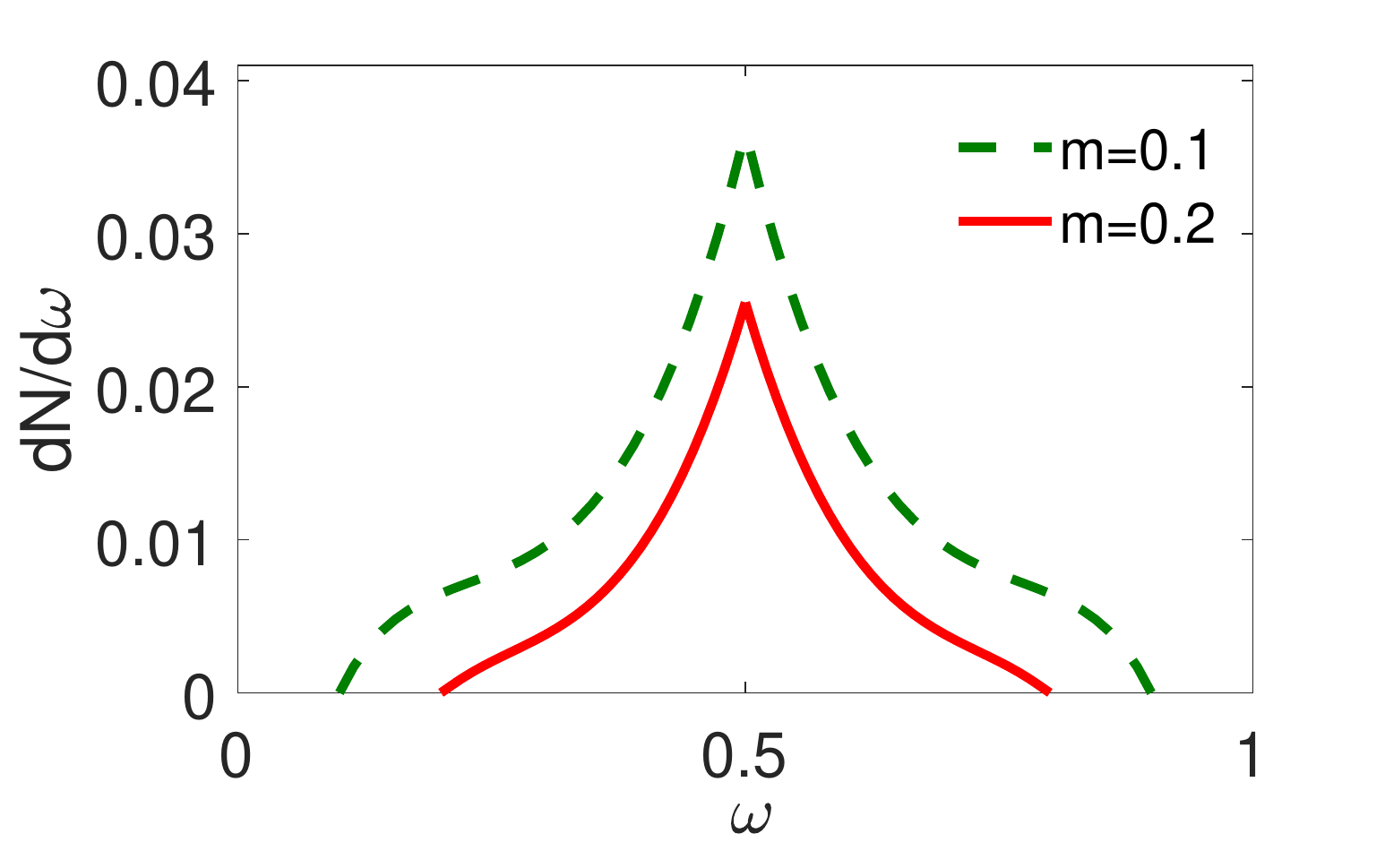}
\caption{The energy frequency of vacuum radiation, where the prefactor $\frac{TS}{4\pi^{2}}$ and driving amplitude $q_{0}$ are set to be $1$.}
\label{fig:freq}
\end{figure}

        This also returns us to the result of the massless field with $m\to 0$. The frequency spectra of vacuum radiation are shown in Figure~\ref{fig:freq}, where the frequency dependence of the vacuum radiation varies with particle mass $m$. The suppression of vacuum radiation by mass can also be inferred from Figure~\ref{fig:freq}.

        Note that the method and results in this section can be easily generalized to $(d+1)$ dimensions.

\section{Conclusions and Discussion}
        In this paper we studied the vacuum radiation of a massive field with one moving mirror. The vacuum friction force on the perfect reflective flat moving mirror in $(1+1)$ dimensions and the spatial spectra of emitted particles in $(3+1)$ dimension are derived explicitly in the non-relativistic limit. The pair emission phenomenon is revealed by means of the canonical quantization procedure and the negative sea picture. We demonstrate that the particle's mass modifies the behavior of vacuum radiation in various aspects. For one, the vacuum emission is suppressed by intrinsic mass as an overall effect. Furthermore, the influence on the spectra of the vacuum radiation power and the vacuum friction force occurs mostly in the low-frequency regime. The spatial spectra of vacuum radiation are also discussed, wherein the particles are emitted at a narrower solid angle when the particle mass grows. Finally, we point out that the results obtained in this article in the limit of $m\to 0$ recover the existing results obtained for the massless field.

        In quantum field theory, the fields modeling the material world interact with each other. As a consequence, the physical quantities, such as mass, charge, and spin, of particles comes from two origins: an intrinsic one from bare particles and an additional one from their interaction. It is shown in \cite{D96,D101int,a,b} that the dressing effect can be very different when the vacuum is undergoing non-adiabatic perturbation. The next steps from this work will be taking the interaction of the fields into consideration and going beyond the non-relativistic regime to study the UV behavior of the vacuum radiation, which may shed light on black hole thermodynamics.

\section*{Acknowlegements}
        We thank J. F. Zhao for discussions of the Green function. We also thank R. Zhang, Y. X. Liu, and K. F. Lyu for help and useful discussion, and we thank M. Moosa for helpful opinions and suggestions.



\begin{thebibliography}{99}



    \bibitem{Itykson} Itzykson, C.; Zuber, J.-B. \textit{Quantum Field Theory};  McGraw Hill: {New York, NY, }USA, 1980; pp. 110--111.


    \bibitem{Casimir48} Casimir, H.B.G. On the attraction between two perfectly conducting plates. {\em Proc. Kon. Ned. Akad. Wet.} {\bf 1948}, {\em 51}, 793.

    \bibitem{Moore} Moore, G.T. Quantum theory of electromagnetic field in a variable-length one-dimensional cavity. {\em J. Math. Phys.} {\bf 1970}, {\em 11}, 2679.

    \bibitem{Davis} Fulling, S.A.; Davies, P.C. W. Radiation from a moving mirror in $2$ dimensional space-time - cnformal anomaly. {\em Proc. R. Soc. A} {\bf 1976}, {\em 348}, 393--414.

    \bibitem{RevofMod} Nation, P.D.; Johansson, J.R.; Blencowe, M.P.; Nori, F. Colloquium: Stimulating uncertainty: Amplifying the quantum vacuum with superconducting circuits. {\em Rev. Mod. Phys.} {\bf 2012}, {\em 1}, 1--24.

    \bibitem{HR1} Hawking, S.W. Particle creation by black-holes. {\em Commun. Math. Phys.} {\bf 1975}, {\em 43}, 199--220.

    \bibitem{HR2} Fabbri, A.; Navarro-Salas, J. \textit{Modeling Black Hole Evaporation};  Imperial College Press: London, UK, 2005; pp. 73--119.

    \bibitem{A57} Schützhold, R.; Plunien, G.; Soff, G. Trembling cavities in the canonical approach. {\em Phys. Rev. A} {\bf 1998}, {\em 57}, 2311--2318.

    \bibitem{Dodonov} Dodonov, V.V.; Klimov A.B.; Man'ko, V.I. Generation of squeezed states in a resonator with a moving wall. {\em Phys. Lett. A} {\bf 1990}, {\em 149}, 225--228.

    \bibitem{A94} Maia Neto, P.A.; Machado, L.A.S. Quantum radiation generated by a moving mirror in free space. {\em Phys. Rev. A} {\bf 1996}, {\em 54}, 3420--3427.

    \bibitem{Neto94} Maia Neto, P.A. Vacuum radiation pressure on moving mirrors. {\em J. Phys. A Math. Gen.} {\bf 1994}, {\em 27}, 2167--2180.

    \bibitem{D82} Ford, L.H.; Vilenkin, A. Quantum radiation by moving mirrors. {\em Phys. Rev. D} {\bf 1982}, {\em 25}, 2569--2575.

    \bibitem{Kardar13} Maghrebi, M.F.; Golestanian, R.; Kardar, M. Scattering approach to the dynamical Casimir effect. {\em Phys. Rev. D} {\bf 2013}, {\em 87}, 025016.

    \bibitem{UE} Unruh, W.G. Notes on black-hole evaporation. {\em Phys. Rev. D} {\bf 1976}, {\em 14}, 870--892.

    \bibitem{lust} Lüst, D.; Vleeshouwers, W. \textit{Black Hole Information and Thermodynamics}; Springer International Publishing:  {Berlin/Heidelberg, Germany,} 
 2019; pp. 9--12.

    \bibitem{BH1} Haro, J.; Elizalde, E. Black hole collapse simulated by vacuum fluctuations with a moving semitransparent mirror. {\em Phys. Rev. D} {\bf 2008}, {\em 77}, 045011.

    \bibitem{BH2} Birrell, N.D.; Davies, P.C.W. \textit{Quantum Fields in Curved Space}; Cambridge University Press: Cambridge, UK, 1984; pp. 102--118.

    \bibitem{BH3} Carlitz, R.D.; Willey, R.S. Lifetime of a black-hole. {\em Phys. Rev. D} {\bf 1987}, {\em 36}, 2336--2341.

    \bibitem{BH4} Carlitz, R.D.; Willey, R.S. Reflections on moving mirrors. {\em Phys. Rev. D} {\bf 1987}, {\em 36}, 2327--2335.

    \bibitem{Unruh92} Wang, Q.D.; Unruh, W.G. Motion of a mirror under infinitely fluctuating quantum vacuum stress. {\em Phys. Rev. D} {\bf 2015}, {\em 92}, 063520.

        \bibitem{D76} Fosco, C.D.; Lombardo, F.C.; Mazzitelli, F.D. Quantum dissipative effects in moving mirrors: A functional approach. {\em Phys. Rev. D} {\bf 2007}, {em 76}, 085007.

    \bibitem{D101} Fosco, C.D.; Junior, D.R.; Oxman, L.E. Quantum effects due to a moving Dirichlet point. {\em Phys. Rev. D} {\bf 2020}, {\em 101}, 065014.

    \bibitem{haro2007} Haro, J. Dynamical Casimir Effect for Scalar Fields I (Particle Creation). {\em Int. J. Theor. Phys.} {\bf 2007}, {\em 46}, 1003--1019.



    \bibitem{D91} Farías, M.B.; Fosco, C.D.; Lombardo, F.C.; Mazzitelli, F.D.; López, A.E.R. Functional approach to quantum friction: Effective action and dissipative force. {\em Phys. Rev. D} {\bf 2015}, {\em 91}, 105020.


    \bibitem{D96} Akhmedov, E.T.; Alexeev, S.O. Dynamical Casimir effect and loop corrections. {\em Phys. Rev. D} {\bf 2017}, {\em 96}, 065001.

    \bibitem{D101int} Akhmedov, E.T.; Lanina, E.N.; Trunin, D.A. Quantization in background scalar fields. {\em Phys. Rev. D} {\bf 2020}, {\em 101}, 025005.

    \bibitem{a} Akopyan, L.A.; Trunin, D.A. Dynamical Casimir effect in nonlinear vibrating cavities. {\em arXiv} {\bf 2021}, arXiv:2012.02129.

    \bibitem{b} Trunin, D.A. Nonlinear dynamical Casimir effect at weak nonstationarity. {\em arXiv} {\bf 2021}, arXiv:2108.07747.


\end{thebibliography}
\end{document}